\documentclass[journal]{IEEEtran}

\usepackage{microtype}
\usepackage{graphicx}
\usepackage{booktabs} 
\usepackage{multirow}
\usepackage{amsmath,amssymb}
\usepackage{enumerate}
\usepackage{subcaption}
\usepackage[linesnumbered,ruled,vlined]{algorithm2e}
\usepackage{algorithmic}
\usepackage[hidelinks]{hyperref}
\usepackage{mathrsfs}
\usepackage{dsfont}
\usepackage{bbm}

\usepackage[table]{xcolor}

\usepackage{todonotes}

\usepackage{cite}
\usepackage{amsthm} 

\usepackage{nicefrac}       

\usepackage[obeyspaces]{xurl}

\usepackage[capitalise]{cleveref}
\crefname{equation}{eq.}{eqs.}

\usepackage{tabularx}
\newcolumntype{Y}{>{\centering\arraybackslash}X} 

\newcommand{\ProbF}{\textsf{P}} 
\newcommand{\Det}{\textsf{D}}   
\newcommand{\fcast}[3]{(\ell^{#1},\,w^{#2},\,s^{#3})}

\DeclareMathAlphabet{\mathdutchcal}{U}{dutchcal}{m}{n}
\SetMathAlphabet{\mathdutchcal}{bold}{U}{dutchcal}{b}{n}
\DeclareMathAlphabet{\mathdutchbcal}{U}{dutchcal}{b}{n}

\newcommand{\netload}{\mathdutchcal{n}}

\newcommand{\Prob}[1]{\sP\left\{{#1}\right\}}
\newcommand{\Risk}[1]{\cR\left\{{#1}\right\}}

\definecolor{dartmouthgreen}{rgb}{0.05, 0.5, 0.06}
\definecolor{bittersweet}{rgb}{1.0, 0.44, 0.37}
\definecolor{brightmaroon}{rgb}{0.76, 0.13, 0.28}
\definecolor{bluegray}{rgb}{0.4, 0.6, 0.8}
\definecolor{brandeisblue}{rgb}{0.0, 0.44, 1.0}
\definecolor{battleshipgrey}{rgb}{0.52, 0.52, 0.51}
\definecolor{cadetgrey}{rgb}{0.57, 0.64, 0.69}
\definecolor{lightgray}{rgb}{0.83, 0.83, 0.83}

\usepackage{amsmath,amsfonts,bm}

\def\eqref#1{equation~\ref{#1}}

\def\1{\bm{1}}

\DeclareMathAlphabet{\mathsfit}{\encodingdefault}{\sfdefault}{m}{sl}
\SetMathAlphabet{\mathsfit}{bold}{\encodingdefault}{\sfdefault}{bx}{n}

\def\sP{{\mathbb{P}}}

\def\sR{{\mathbb{R}}}

\def\cD{\mathcal{D}}

\def\cR{\mathcal{R}}
\def\cT{\mathcal{T}}

\def\cV{\mathcal{V}}

\newcommand{\E}{\mathbb{E}}

\newif\ifarxiv
\arxivtrue

\usepackage{comment}

\begin{document}

\title{Source-Agnostic Sizing of Flexibility Reserves}

\author{Napoleon~Costilla-Enriquez,~\IEEEmembership{Member,~IEEE,}
    ~Miguel~A.~Ortega-Vazquez,~\IEEEmembership{Senior Member,~IEEE,}
    ~Aidan~Tuohy,~\IEEEmembership{Member,~IEEE,}
    ~Erik~Ela,~\IEEEmembership{Member,~IEEE}

    \thanks{
        N. Costilla-Enriquez, M. A. Ortega-Vazquez, A. Tuohy, and E. Ela are with the Electric Power Research Institute, Palo Alto, California, \{ncostilla-enriquez, maov, atuohy, eela\}@epri.com.}%
}

\ifarxiv
    \onecolumn
    \thispagestyle{empty}
    \noindent
    This work has been submitted to the IEEE for possible publication.
    Copyright may be transferred without notice, after which this version
    may no longer be accessible.
    \vspace*{\fill}
    \setcounter{page}{0}
    \twocolumn
\fi

\maketitle

\begin{abstract}
      Growing shares of variable renewable energy sources (VRES) increase forecast uncertainty and variability, requiring flexibility reserves that adapt to changing operating conditions and reflect the likelihood and severity of forecast deviations.  Conventional fixed-rule or Gaussian approaches misrepresent asymmetric, heavy-tailed errors, leading to inefficient procurement or optimistic risk estimates.  This paper presents a source-agnostic framework that constructs conditional error distributions for load, wind, and solar from historical deviations or probabilistic forecasts, combines them into a conditional net-load error distribution, and derives upward and downward reserves using coverage- and risk-based criteria.  Nonparametric density estimation captures empirical error behavior without a parametric assumption, while a Conditional Value-at-Risk (CVaR)-based metric quantifies expected uncovered deviations.  Experiments on a New York Independent System Operator (NYISO)-based synthetic dataset show that the framework meets target coverage with lower reserve volumes than static benchmarks and avoids the tail-risk distortion of Gaussian-based methods.  Because reserves are built directly from the resource uncertainty distributions, the framework is transparent and interpretable, and it integrates seamlessly into existing production-cost models as deterministic reserve constraints, avoiding scenario-based stochastic optimization. It is implemented in the Electric Power Research Institute's (EPRI) DynADOR tool for operational reserve scheduling.
\end{abstract}


\section{Introduction}
Many power systems now operate with growing shares of variable renewable energy sources (VRES). The output of these resources is weather-driven, increasing variability and forecast uncertainty in system operations. At the same time, electricity demand is changing due to transportation electrification and the integration of new large loads, such as data centers. As a result, system operators must schedule sufficient operating reserves\footnote{In this work, operating reserve is defined as the headroom and footroom held in a scheduling process, e.g., day-ahead, and released in a subsequent process, e.g., real-time, to accommodate deviations from forecasted quantities.} to maintain balance between supply and demand under uncertain operating conditions.

Existing methods for determining operating reserves are commonly based on rules of thumb or statistical assessments of forecast errors around a single projected trajectory of stochastic variables such as load and VRES output (e.g.,~\cite{bouffard2011value,dvorkin2014assessing,ortega2008estimating,ela2011operating,holttinen2012methodologies,bapin2018estimation,bruninx2014statistical, OV-K_2009}).
While these methods are simple to implement and widely used in practice, they have important limitations.

First, rules of thumb often estimate reserve levels from aggregated historical data and provide little or no conditioning on the current operating state. Other methods condition reserves on variables such as load, wind production, solar production, or net load, but still infer uncertainty from historical deviations around deterministic forecasts. Consequently, they do not use the forecast-specific information contained in probabilistic forecasts, such as the expected spread, skewness, or tail behavior for the operating day. Second, parametric methods, especially those that assume Gaussian forecast errors, impose symmetry and light tails on the error distribution \cite{holttinen2012methodologies,krad2015quantifying}. These assumptions contradict empirical evidence, since real forecast errors are often skewed and heavy-tailed, particularly under extreme weather events~\cite{bruninx2014statistical,hodge2011distribution}. Third, coverage-based reserve sizing controls the frequency of reserve violations but does not directly quantify the severity of uncovered deviations.

Recent studies have incorporated probabilistic forecasts into reserve determination~\cite{hong2020energy,NCE2023}, improving adaptability to forecast uncertainty. However, existing approaches do not provide a unified framework that can use either historical deviations or probabilistic forecasts, compare the reserve and risk implications of each information source, and account for both the probability and magnitude of uncovered deviations. In practice, the size of a reserve shortfall is as important as its probability, particularly in systems with increasing renewable penetration.

To address these limitations, this paper proposes a source-agnostic framework for flexibility reserve sizing that can use either historical forecast deviations or probabilistic forecasts as uncertainty inputs. The main contributions are:
\begin{enumerate}
      \item \textit{Dynamic reserve sizing}: The proposed framework adapts reserve requirements to the prevailing system state using conditional probability models based on operating conditions.
      \item \textit{Source-agnostic uncertainty modeling}: Conditional error distributions can be constructed from historical forecast deviations or directly from probabilistic forecasts, allowing both information sources to be used within the same reserve-sizing framework.
      \item \textit{Nonparametric conditional densities}: For deterministic forecasts, $k$-nearest-neighbor density estimation is used to construct empirical error distributions without imposing Gaussian assumptions, preserving skewness and heavy-tail behavior in load, wind, and solar deviations.
      \item \textit{Risk-aware reserve sizing}: In addition to coverage-based requirements, the framework uses a Conditional Value-at-Risk (CVaR)-based metric to quantify expected uncovered deviations and distinguish upward and downward risk.
      \item \textit{Transparency and interpretability}: Reserve requirements are constructed directly from the uncertainty distributions of load, wind, and solar, so every reserve value can be traced back to the underlying error distributions and conditioning variables. This transparent construction lets operators understand the rationale behind a reserve value, increasing confidence and supporting defensible reserve decisions.
      \item \textit{Practical deployment and seamless integration}: The proposed models are implemented in EPRI's DynADOR tool \cite{dynador}, supporting reserve scheduling in real world operational settings.  Because the framework produces explicit upward and downward reserve requirements, it integrates directly into existing production-cost models as deterministic reserve constraints, aligning with current operator practice rather than requiring a scenario-based stochastic optimization and its added computational complexity.
\end{enumerate}

Using an NYISO-based synthetic dataset, we show that the proposed framework improves reserve-volume efficiency relative to static reserve benchmarks while maintaining reliability. It avoids the tail-risk distortion caused by Gaussian assumptions and captures directional differences between upward and downward reserve needs. The experiments also compare deterministic and probabilistic forecast inputs, showing how probabilistic information affects day-specific reserve sizing and risk assessment.

The paper is organized as follows. \cref{sec:background} presents the problem formulation. \cref{sec:approach} introduces the source-agnostic reserve sizing framework. \cref{sec:experiments} presents the numerical experiments, and \cref{sec:conclusion} concludes the paper.

\section{Problem Formulation}\label{sec:background}
To maintain balance between supply and demand in power systems, operators must size upward and downward flexibility reserves to account for deviations between forecasted and actual net load. This task is increasingly challenging due to two fundamental phenomena: \textit{uncertainty} and \textit{variability} \cite{ortega2022generation}.
\begin{itemize}
    \item \textit{Uncertainty} captures the limitations in forecasting future system conditions, such as electricity demand or renewable generation output. For example, even with advanced forecasting tools, it remains challenging to predict solar irradiance or wind speed hours ahead precisely. This uncertainty leads to potential mismatches between planned and actual system conditions, requiring reserves to buffer against forecast errors.
    \item \textit{Variability}, in contrast, refers to the intrinsic temporal fluctuations in system inputs, even if forecasts were perfectly accurate. Renewable energy sources such as wind and solar undergo continuous changes in output due to natural weather dynamics. A passing cloud or a gust of wind can lead to rapid changes in generation, independent of forecast accuracy. These short-term, high-frequency fluctuations require flexible, responsive reserves to maintain frequency stability.
\end{itemize}
The integration of VRES exacerbates both uncertainty and variability. Unlike conventional power plants, which are dispatchable and controllable, VRES are weather-dependent and variable, introducing additional complexity into balancing operations. As a result, modern power systems must utilize more sophisticated operational strategies and more flexible resources to ensure reliable system performance under increasingly dynamic and uncertain conditions.

\begin{figure}[ht]
    \centering
    \includegraphics[width=2.5in]{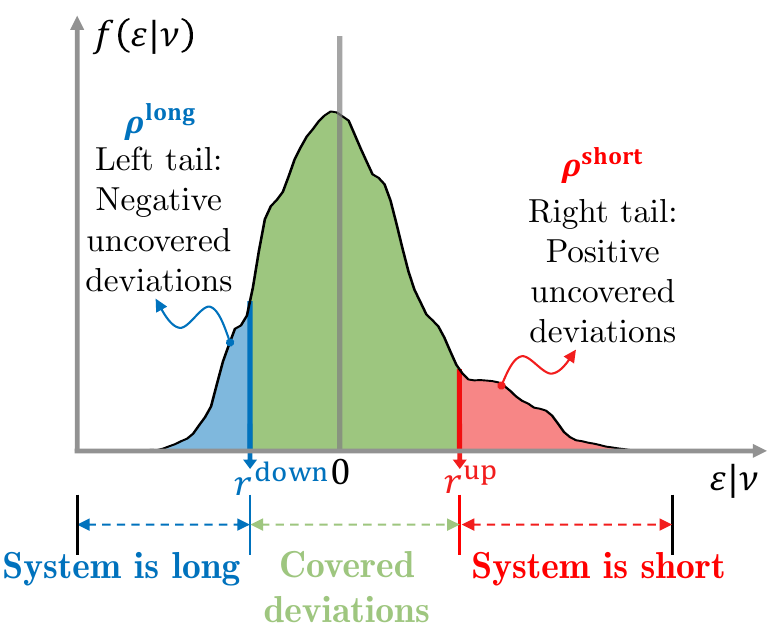}
    \caption{System deviations with respect to allocated reserves.}
    \label{fig:risk_diagram}
\end{figure}

Commitment and dispatch processes must schedule generation to meet forecasted system-wide net demand (i.e., load minus wind and solar power generation), while ensuring the system has sufficient reserves to accommodate forecast deviations due to variability and uncertainty.
The relationship between the forecasted and actual net demand is given as:
\begin{equation}\label{eq:net_load_F_error}
    \netload^{\textrm{A}}_t = \netload^{\textrm{F}}_t + \varepsilon_t,
\end{equation}
where $\netload^{\textrm{F}}_t$ is the (e.g., day-ahead) net load forecast value, $\netload^{\textrm{A}}_t$ is the actual realization (e.g., real-time), and $\varepsilon_t$ is the forecast error/deviation.
The term $\varepsilon_t$ is a random variable with an unknown probability density function (PDF), driven by exogenous factors such as weather conditions (e.g., wind speed, solar irradiance, temperature) and electricity consumption patterns.
A value $\varepsilon_t > 0$ indicates reserve needs in the upward direction $r^{\textrm{up}}\in\sR_{\geq0}$, and $\varepsilon_t < 0$ indicates reserve needs in the downward direction $r^{\textrm{down}}\in\sR_{\geq0}$, as illustrated in \cref{fig:risk_diagram}. To ensure system balance under any realization of $\varepsilon_t$, it is essential to establish an accurate reserve sizing model.

In general, $\varepsilon_t$ may reflect forecast uncertainty, variability, or both. The proposed framework can accommodate any characterization of errors or any combination of them. Throughout this work, we focus on uncertainty-driven deviations to emphasize the method development and the framework's core contribution. Still, the method is general to account for deviations from both uncertainty and variability.
%

\section{Proposed Method}\label{sec:approach}
\begin{figure}[ht]
     \centering
     \includegraphics[width=3.2in]{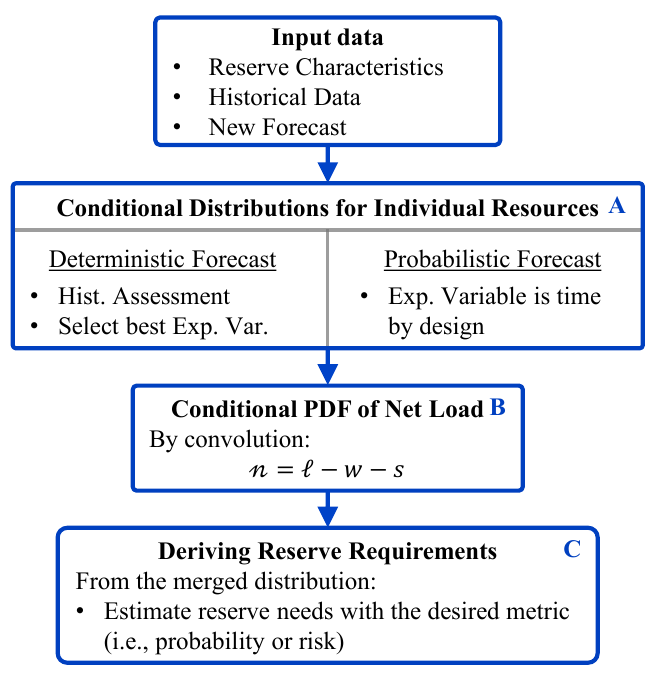}
     \caption{Diagram of the proposed method.}
     \label{fig:method_diagram}
\end{figure}

The proposed method sizes flexibility reserves by integrating both deterministic and probabilistic load and VRES forecasts to construct conditional probability distributions of net load errors and determine upward and downward reserve requirements. The workflow is illustrated in \cref{fig:method_diagram}.
The objective is to obtain dynamic reserve requirements that reflect the system's operating conditions. The process consists of: (i) constructing conditional distributions for load, wind, and solar forecast errors, (ii) merging these into a conditional distribution of net load uncertainty, and (iii) deriving reserve requirements through probabilistic and risk-based metrics.
By construction, every reserve value is obtained from these explicitly modeled distributions and their conditioning variables, so each step is transparent and interpretable, letting operators trace a reserve requirement back to the resource-level uncertainty that produced it.
This section defines the modeling components used in the framework; their empirical effect is evaluated in \cref{sec:experiments}, where candidate deterministic configurations are compared and the selected one is benchmarked against static and Gaussian-based methods.
%

\subsection{Conditional Distributions for Individual Resources (Load, Wind, Solar)}
\begin{figure}[ht]
     \centering
     \includegraphics[width=2.8in]{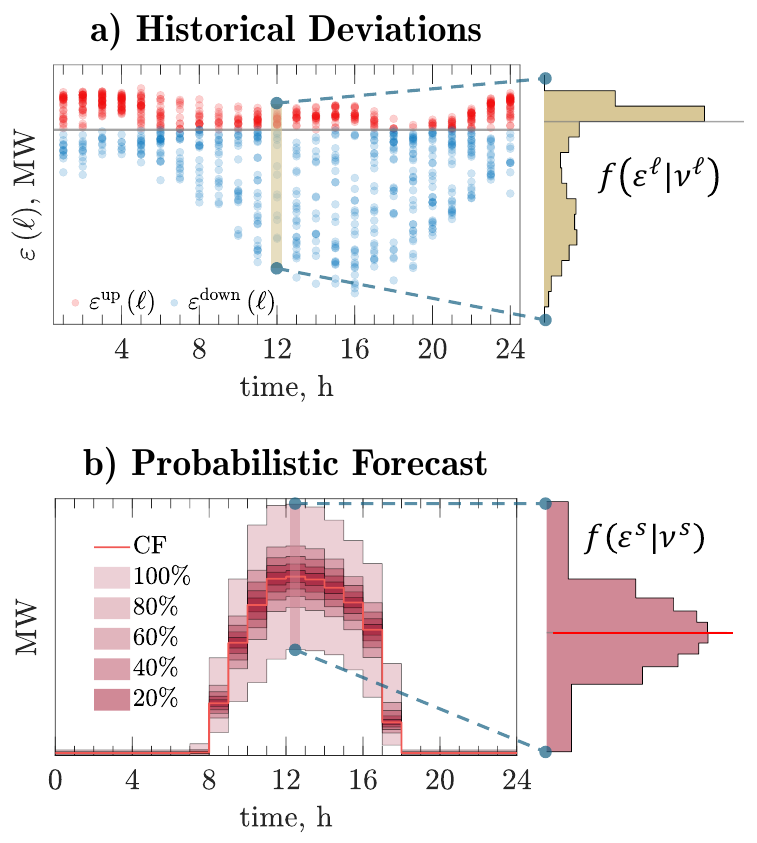}
     \caption{Conditional distributions from a) historical deviations and b) probabilistic forecasts}
     \label{fig:historical_vs_prob_diagram}
\end{figure}

The conditional probability distributions of forecast errors of load, wind, and solar for either (i) deterministic forecasts with historical deviations or (ii) probabilistic forecasts are computed as illustrated in \cref{fig:historical_vs_prob_diagram}.
%

\subsubsection{Conditional PDF from Deterministic Forecasts}\label{sec:historical_data}
Given $H$ historical pairs of forecasted and realized values $\left\{ \left( X^{\textrm{F}}_i, X^{\textrm{A}}_i \right) \right\}_{i=1}^H$, historical deviations are computed as the difference between actual and forecasted values.
In practice, however, these two sources often have different temporal resolutions (e.g., hourly day-ahead forecasts vs. 5-minute measurements), so the actual series must first be downsampled to the forecast resolution before computing the errors.
The historical deviation for resource $X$ is then calculated as
\begin{equation}\label{eq:historical_error}
     \varepsilon^{X}_i = X^{\textrm{A}}_i - X^{\textrm{F}}_i, \qquad i=1,2\ldots,H,
\end{equation}
where $X^{\mathrm{F}}_i$ and $X^{\mathrm{A}}_i$ denote the forecasted and downsampled actual quantities, respectively.
Based on these historical errors, our target is to construct the conditional distribution given an explanatory variable $\nu^{X}$: $f\left( \varepsilon^{X} \;|\; \nu^{X} \right)$.

\paragraph{Step 1: Define candidate explanatory variables}
From the historical forecast time series, a set of candidate explanatory variables can be constructed to capture the dependency structure of forecast errors. Examples include the forecasted production level, its rate of change and absolute rate of change, and temporal indicators such as hour of day, day of week, or season. However, only variables physically or operationally related to forecast errors should be considered, since others may exhibit spurious dependence with no meaningful explanatory power. In other words, \emph{statistical dependence does not imply causation}, and explanatory-variable selection must be guided by domain knowledge and data analysis. Based on these considerations, we assemble a candidate set of explanatory variables as
\begin{equation}
     \cV
     =
     \left\{
     X^{\textrm{F}},\; \Delta X^{\textrm{F}},\; |\Delta X^{\textrm{F}}|,\; t^{\textrm{F}}
     \right\}
\end{equation}
where $X^{\textrm{F}}$ is the forecasted resource level, $\Delta X^{\textrm{F}}$ is its one-step change, and $t^{\textrm{F}}$ represents temporal information from which calendar indicators can be derived.

\begin{figure}[htb]
     \centering
     \begin{subfigure}[t]{1.65in}
          \centering
          \includegraphics[width=\textwidth]{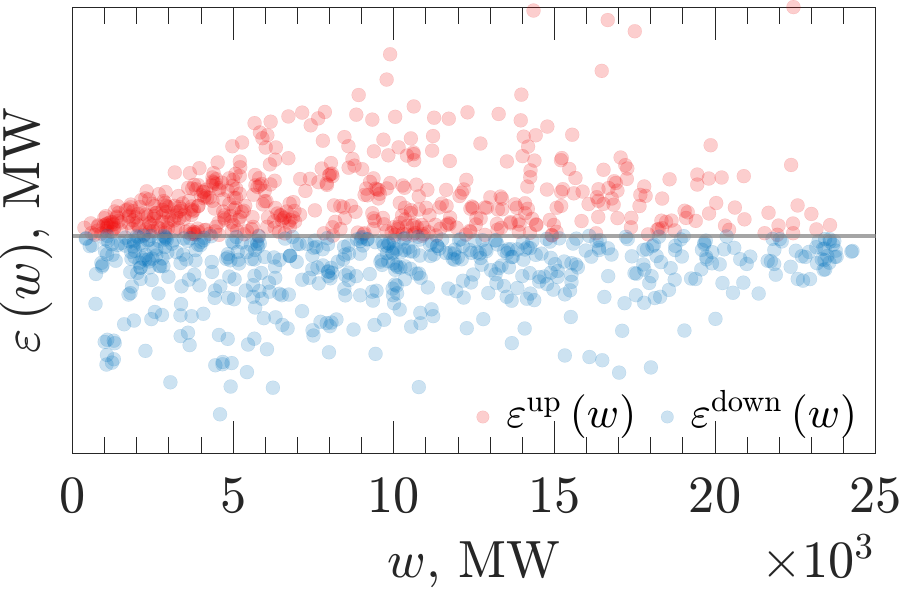}
          \caption{Production.}
     \end{subfigure}
     \hfill
     \begin{subfigure}[t]{1.65in}
          \centering
          \includegraphics[width=\textwidth]{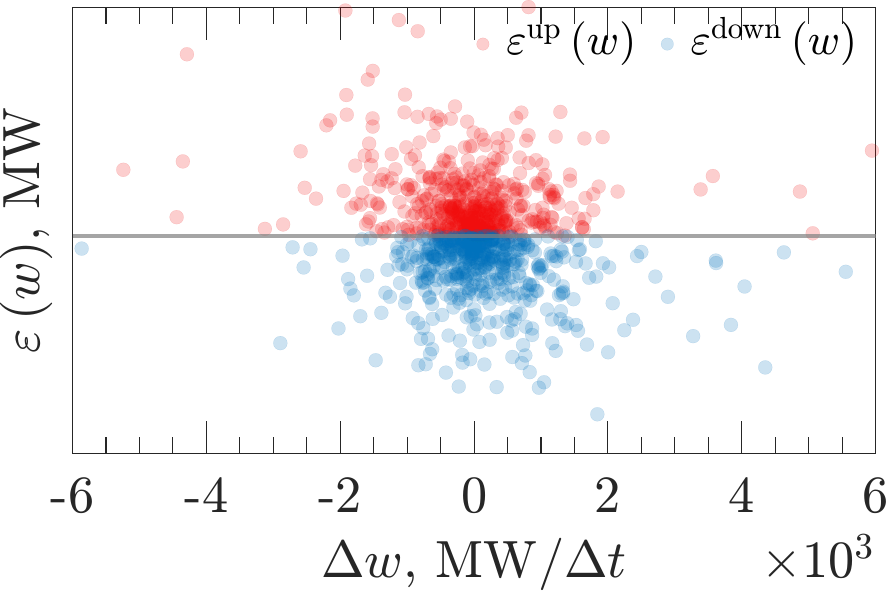}
          \caption{Rate of change.}
     \end{subfigure}
     \hfill
     \begin{subfigure}[t]{1.65in}
          \centering
          \includegraphics[width=\textwidth]{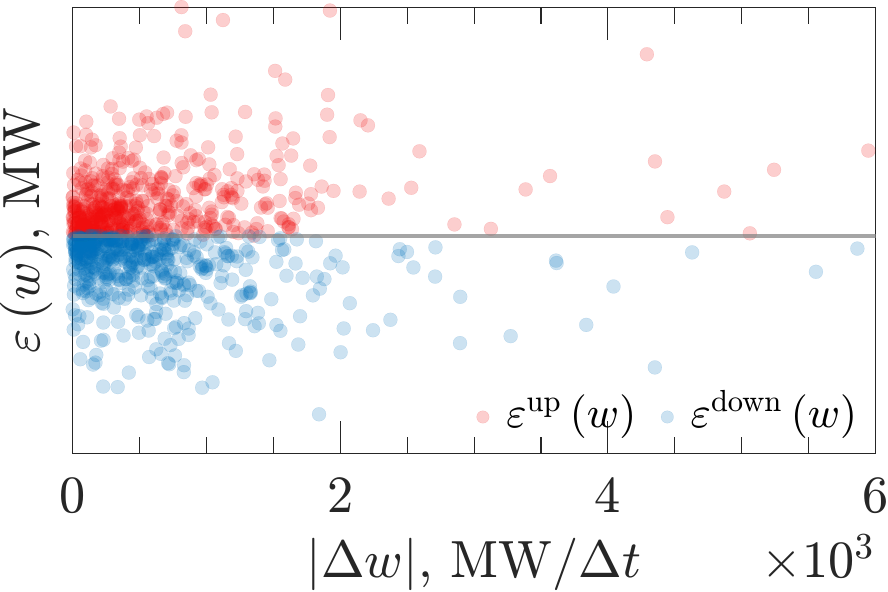}
          \caption{Absolute value of rate of change.}
     \end{subfigure}
     \hfill
     \begin{subfigure}[t]{1.65in}
          \centering
          \includegraphics[width=\textwidth]{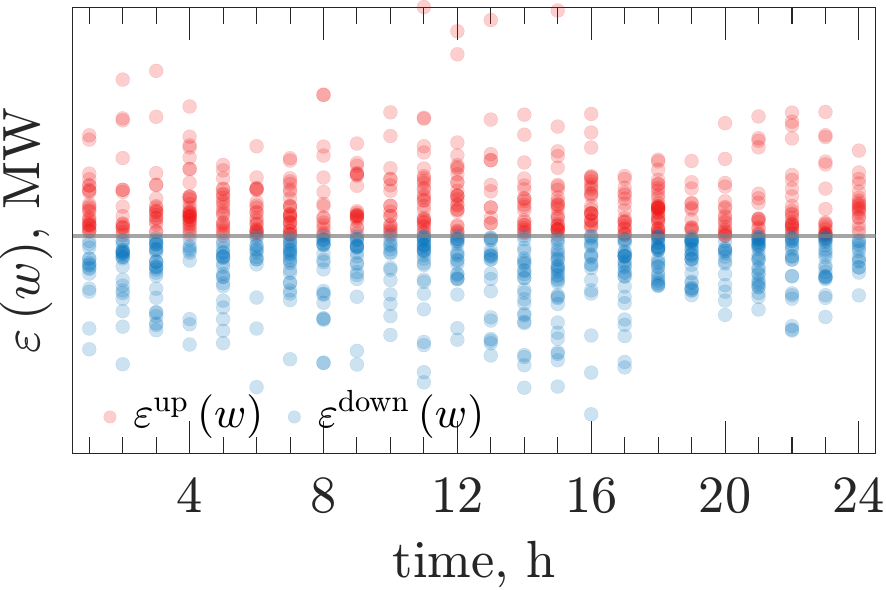}
          \caption{Time.}
     \end{subfigure}
     %
     \caption{Comparison of different explanatory variables.}
     \label{fig:explanatory_variables}
\end{figure}

\paragraph{Step 2: Explanatory variable selection}
\cref{fig:explanatory_variables} shows the joint historical distribution of wind forecast errors in relation to the candidate explanatory variables for the NYISO dataset.  The distinct patterns in the scatter plots suggest that each variable captures the error behavior in different ways. Consequently, the choice of explanatory variable is critical, as it directly influences the strength and nature of the observed dependency.
%

The next step is to measure the statistical strength between the errors and potential explanatory variables.
Correlation has been widely applied in reserve modeling approaches \cite{holttinen2012methodologies, krad2015quantifying}, and provides a simple measure of linear dependence. Because forecast errors can also depend nonlinearly on operating conditions, we also consider mutual information (MI), which measures the reduction in uncertainty of one variable given another and captures both linear and nonlinear dependencies \cite{cover2006elements}.
To make MI values comparable across datasets and explanatory variables, we instead use the Normalized Mutual Information (NMI) score, which bounds the similarity measure to the interval $\left[0,1\right]$. The Normalized Mutual Information, using the \textit{arithmetic mean normalization}, is expressed as
\begin{equation}
     NMI(X,Y) = \frac{2\, I(X;Y)}{H(X) + H(Y)},
\end{equation}
where $I\left(\cdot\right)$ denotes mutual information and $H\left(\cdot\right)$ denotes entropy. By construction, $0$ indicates independence, while $1$ indicates identical variables.

\begin{figure}[ht]
     \centering
     \includegraphics[width=1.4in]{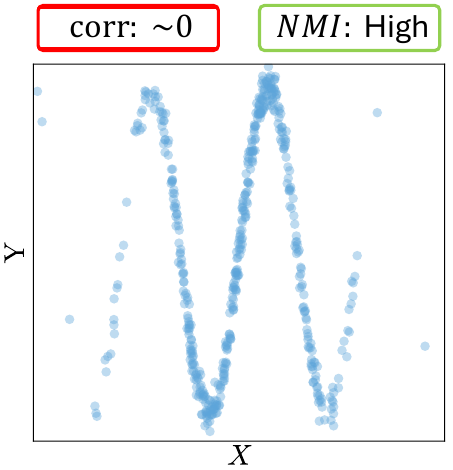}
     \caption{Correlation vs Normalized Mutual Information.}
     \label{fig:corr_vs_MI}
\end{figure}

\cref{fig:corr_vs_MI} provides an example where correlation fails to capture a nonlinear dependency, whereas NMI detects the statistical dependence between the variables. Thus, both correlation and NMI can be used as selection criteria, with NMI offering a more general measure when the relationship is nonlinear.

For each criterion, the explanatory variable with the strongest dependence is selected and used to construct the conditional error distribution. The numerical experiments compare correlation- and NMI-based selection to assess how this modeling choice affects reserve sizing.
%

\paragraph{Step 3: Estimate the conditional density $f(\varepsilon \mid \nu)$}
The conditional error distributions are estimated using nonparametric methods, since they rely directly on the data rather than assuming a fixed functional form. This is useful because forecast errors in load, wind, and solar may exhibit skewed or heavy-tailed densities that are not well represented by simple parametric models.

To construct the conditional density around a \emph{forecasted value}, we focus on samples in its vicinity and estimate the local distribution. Among nonparametric techniques, the proposed deterministic configuration uses the \textit{nearest-neighbor} ($k$-NN) density estimator. This method evaluates density at a point based on the distance to its $k$ nearest data points, dynamically adjusting the neighborhood size to reflect local data concentration \cite{bishop2006pattern}.

The $k$-NN estimator adapts locally: the estimated density contracts in regions of high data density and expands in areas of low data density. This property is relevant for reserve sizing because the dispersion of forecast errors varies with operating conditions. Accordingly, the conditional densities $f(\varepsilon \mid \nu)$ are derived directly from the data, centered on the forecasted value, and estimated via the $k$-NN method. Alternative binning choices are evaluated in \cref{sec:ablation_study}.

\subsubsection{From Probabilistic Forecasts}\label{sec:prob_forecast}
Probabilistic forecasts directly provide the conditional distribution of a variable for a given time horizon, as illustrated in \cref{fig:historical_vs_prob_diagram}b. Unlike the deterministic case, where explanatory variables must be identified from historical data, the conditioning variable here is already specified by design: time. Each forecast corresponds to a probability distribution of possible load, wind, or solar realizations at a given time, constructed under the prevailing weather conditions (e.g., expected cloud cover or wind regime for that day and hour).

This means that the forecast itself delivers the information required to build the conditional distribution $f(\varepsilon \mid \nu)$.  The probabilistic forecast may be represented in different but equivalent forms (e.g., quantiles, probabilities of exceedance, or prediction intervals). For implementation, we rely on the cumulative distribution function (CDF), from which the probability density function (PDF) can be derived when needed.

In our framework, these probabilistic distributions are integrated directly into the reserve sizing process.  They capture day-specific characteristics that historical records alone may not reflect, allowing the conditional error distribution to be tailored to the unique forecasted operating conditions \cite{NCE2023}.

\subsection{Conditional PDF of Net Load Fluctuations}\label{sec:cond_pdf_netload}
The previous subsections described how conditional error distributions are obtained for each resource (load, wind, and solar) using either historical deviations or probabilistic forecasts. The next step is to combine these into a single conditional distribution of net load fluctuations, which is required to estimate the system's reserve requirements.

Formally, the net load error $\varepsilon$ given the predictor variables $\nu$ can be expressed as the combination of load, wind, and solar forecast errors
\begin{equation}\label{eq:net_load_rv_cond}
     \varepsilon \;|\; \nu
     =
     \varepsilon \;|\; \nu^{\ell}, \nu^{w}, \nu^{s}
     =
     \varepsilon^{\ell} \;|\; \nu^{\ell} -
     \varepsilon^{w} \;|\; \nu^{w} -
     \varepsilon^{s} \;|\; \nu^{s}
     ,
\end{equation}
where $\nu^{\ell}_t$, $\nu^{w}_t$, and $\nu^{s}_t$ are the explanatory variables (or conditioning information) for load, wind, and solar, respectively.

To obtain the conditional PDF of the net load, the individual error distributions must be merged. Since the errors are modeled without assuming a fixed parametric form, we employ convolution, denoted by `$*$', which provides the distribution of the sum of independent random variables \cite{shynk2012probability}.
Applying this to \cref{eq:net_load_rv_cond}, the conditional net load distribution is given by
\begin{equation}\label{eq:net_load_PDF}
     f\left( \varepsilon \;\middle|\; \nu \right)
     =
     f\left( \varepsilon^{\ell} \;|\; \nu^{\ell} \right)
     *
     f\left( -\varepsilon^{w} \;|\; \nu^{w} \right)
     *
     f\left( -\varepsilon^{s} \;|\; \nu^{s} \right)
     .
\end{equation}

In practice, the individual PDFs $\varepsilon^{\ell} \;|\; \nu^{\ell}$, $\varepsilon^{w} \;|\; \nu^{w}$, and $\varepsilon^{s} \;|\; \nu^{s}$ may come from historical forecast errors or directly from probabilistic forecasts, depending on the available data (see \cref{fig:historical_vs_prob_diagram}). The convolution step integrates this heterogeneous information into a unified conditional distribution of net load fluctuations.

Note that in this case, it is assumed that the errors in load, wind, and solar forecasts are independent. This assumption is supported by both theory and empirical analysis of the NYISO dataset: load errors arise mainly from demand-side behavior, while weather dynamics drive wind and solar errors. Correlation coefficients among these errors are consistently close to zero, justifying the independence approximation; the supporting results are presented in \cref{tab:RVs-dependencies}. While independence enables a tractable convolution, future work could relax it when material dependencies are present, such as those arising from spatially correlated weather or neighboring-area conditions.

\subsection{Deriving Reserve Requirements}
Once the conditional distribution of net load fluctuations is obtained (\cref{sec:cond_pdf_netload}), the next step is to translate this distribution into reserve requirements. In practice, this means constructing prediction intervals that guarantee sufficient upward and downward reserves while controlling a specific metric (e.g., probability or risk).

Let $\left\{(\nu_h,\varepsilon_h)\right\}_{h=1}^H$ denote the training samples of explanatory variables and forecast errors, and let $\varepsilon_{H+1}$ be the error associated with the next forecast.
While these pairs are often drawn from historical records, in our framework, they may also come from any constructed conditional distribution, whether derived from deterministic forecasts (via explanatory variables and density estimation) or from probabilistic forecasts that already provide error distributions.
The objective is to build a prediction interval $C(\nu_{H+1})$ that covers $\varepsilon_{H+1}$ with a desired reliability level. Two complementary metrics are considered:
\begin{subequations}\label{eq:confidence_intervals}
     \begin{alignat}{2}
          \label{eq:CI_miscoverage}
          \Prob{\varepsilon_{H+1} \in C_{\alpha}\left( \nu_{H+1} \right)}
           & \geq
          1 - \alpha
          ,       \\
          \label{eq:CI_risk}
          \Risk{
               \varepsilon_{H+1} \notin C_{\rho}\left( \nu_{H+1} \right)
          }
           & \leq
          \gamma^{\textrm{lim}}
          ,
     \end{alignat}
\end{subequations}
where $\alpha$ is the target miscoverage rate and $\gamma^{\textrm{lim}}$ is the acceptable risk tolerance. \Cref{eq:CI_miscoverage} ensures that the interval captures the error with high probability, while \cref{eq:CI_risk} limits the severity of deviations that fall outside the interval.

In this framework, $C_{\alpha}(\cdot)$ is estimated using conditional quantile regression (\cref{sec:conditional_quantile}), and $C_{\rho}(\cdot)$ is obtained through conditional risk measures (\cref{sec:conditional_risk}). Together, these approaches allow reserve requirements to be dimensioned to balance probabilistic coverage with explicit control of risk exposure.

\subsubsection{Conditional Quantile Regression}\label{sec:conditional_quantile}
One way to construct prediction intervals is through conditional quantile regression \cite{koenker1978regression}, which estimates the quantiles of the error distribution conditional on the forecast information. For a given miscoverage rate $\alpha$, we compute the lower and upper conditional quantiles,
$q_{\alpha^{\textrm{low}}}\left(x\right),
     q_{\alpha^{\textrm{high}}}\left(x\right)$,
with $\alpha^{\textrm{low}} = \alpha/2$ and $\alpha^{\textrm{high}} = 1-\alpha/2$. These quantiles define the prediction interval
\begin{equation}
     C\left(x\right) =
     \left[
          q_{\alpha^{\textrm{low}}}\left(x\right),
          q_{\alpha^{\textrm{high}}}\left(x\right)
          \right],
\end{equation}
which by construction ensures
\begin{equation}\label{eq:CI_quantile}
     \Prob{
          Y \in C\left(X\right) \;|\; X=x
     }
     \geq 1 - \alpha
     .
\end{equation}

The interval bounds naturally translate into upward and downward reserve requirements:
\begin{subequations}
     \begin{alignat}{2}
          r^{\textrm{up}}   & =
          \max\left\{ 0, q_{\alpha^{\textrm{high}}}\left(x\right) \right\},
          \\
          r^{\textrm{down}} & =
          \max\left\{ 0, -q_{\alpha^{\textrm{low}}}\left(x\right) \right\}.
     \end{alignat}
\end{subequations}

This approach directly links statistical quantiles of forecast errors with reserve sizing: the system operator specifies the acceptable miscoverage rate $\alpha$, based on reliability targets, historical backtesting, or cost-risk tradeoffs, and the corresponding conditional quantiles determine the reserves needed to achieve that reliability.

\subsubsection{Conditional Risk Measures}\label{sec:conditional_risk}
While prediction intervals describe how often forecast errors are expected to occur within a certain range, they do not say anything about the magnitude of the errors when those limits are exceeded, as illustrated in \cref{fig:risk_diagram}. In power system operations, both aspects matter: a small but frequent deviation may continuously require reserve deployment and pose manageable but continuous stress to the system, while a rare but severe deviation could produce severe imbalances when it materializes and reduce system reliability. For this reason, \emph{risk}, a more comprehensive metric that captures likelihood and magnitude, should be used to assess expected system performance.

Different risk measures exist, and the best choice depends on the operator's priorities. Some systems may aim to reduce the frequency of shortfalls, while others may focus on minimizing their size when they do occur.
A widely used risk metric is the Conditional Value at Risk (CVaR)\cite{rockafellar2000optimization}, which captures the expected shortfall in the tails of the distribution beyond a specified quantile. CVaR is coherent (subadditive) and provides a more complete view of tail risks than variance or Value-at-Risk (VaR). Formally, for a given confidence level $\alpha$, the CVaR of the error distribution is \cite{cornuejols2018optimization}
\begin{equation}\label{eq:CVaR}
     \textrm{CVaR}_\alpha =
     \E
     \left[
          Y
          \;\big|\;
          Y
          \geq q_{\alpha}
          \right]
     =
     \frac{1}{1-\alpha}\int_{\alpha}^{1} q_{\beta}\left(x\right) d\beta
     .
\end{equation}
where $q_{\beta}\left(x\right)$ is the conditional quantile function.
While CVaR captures the average severity of extreme deviations, it does not account for their likelihood.

In our setting, the focus is not on the overall tail magnitude but on the expected size of deviations that exceed the reserves defined by a quantile. To capture this, we introduce
\begin{equation}
     \Delta_\alpha\left(x\right) :=
     \left(
     \textrm{CVaR}_\alpha\left(x\right) - q_\alpha\left(x\right)
     \right)
     \left( 1 - \alpha \right)
     .
\end{equation}

This quantity has a clear mathematical and physical interpretation:
\begin{itemize}
     \item The term $\left(
                \textrm{CVaR}_\alpha\left(x\right) - q_\alpha\left(x\right)
                \right)$ represents the average excess deviation beyond the reserve level $q_\alpha$.
     \item Multiplying by the tail probability $(1-\alpha)$ weights this excess by how often such violations occur.
\end{itemize}

This measure, denoted as $\Delta_{\alpha}(x)$, increases when extreme deviations become either more frequent or more severe, making it directly relevant for \emph{reliability-aware reserve sizing}. It provides a clearer and more operationally meaningful representation of uncovered risk than CVaR alone, telling operators not just how often errors happen, but also how severe they can be when they do.

Using this definition, we construct conditional risk functions for the two directions of imbalance:
\begin{subequations}
     \begin{alignat}{2}
          \rho_{\gamma^{\textrm{long}}}
           & :=
          \inf \left\{
          y \in \sR : \Delta^{\textrm{long}}_{\alpha}
          \left(\cdot\right)
          \geq \gamma^{\textrm{long}}
          \right\}
          ,     \\
          \rho_{\gamma^{\textrm{short}}}
           & :=
          \inf \left\{
          y \in \sR : \Delta^{\textrm{short}}_{\alpha}
          \left(\cdot\right)
          \geq \gamma^{\textrm{short}}
          \right\}
          ,
     \end{alignat}
\end{subequations}
which correspond to downward and upward deviations, respectively. The associated risk thresholds $\gamma^{\textrm{long}}$ and $\gamma^{\textrm{short}}$ define the prediction interval
\begin{equation}
     C_{\rho}\left(x\right) =
     \left[
          \rho_{\gamma^{\textrm{long}}}\left(x\right),
          \rho_{\gamma^{\textrm{short}}}\left(x\right)
          \right]
     ,
\end{equation}
These thresholds can be set separately, since longfall and shortfall events may have different cost and reliability implications.
The resulting interval satisfies
\begin{equation}
     \Risk{
          Y \notin C\left(X\right) \;|\; X=x
     }
     \leq \gamma^{\textrm{lim}},
\end{equation}
with $\gamma^{\textrm{lim}} = \gamma^{\textrm{long}} + \gamma^{\textrm{short}}$.

Finally, upward and downward reserve requirements follow directly as
\begin{subequations}
     \begin{alignat}{2}
          r^{\textrm{up}}   & =
          \max\left\{ 0, \rho_{\gamma^{\textrm{short}}}\left(x\right) \right\},
          \\
          r^{\textrm{down}} & =
          \max\left\{ 0, -\rho_{\gamma^{\textrm{long}}}\left(x\right) \right\}.
     \end{alignat}
\end{subequations}

\section{Numerical Experiments}\label{sec:experiments}

\subsection{Dataset and Experimental Setup}\label{sec:dataset_setup}
The proposed framework is evaluated using an NYISO-based synthetic dataset provided by the Advanced Research Projects Agency–Energy (ARPA–E) under Award Number DE-AR0001276. The dataset represents a high-VRES scenario built using 2019 weather conditions and load profiles, and includes load, wind, and solar generation time series for NYISO's eleven zones (A–K) and the aggregated system ($\Sigma$). Both deterministic and probabilistic forecasts are available, together with synthetic realizations, all sampled at a 60-minute resolution.

Before conducting the main experiments, it is necessary to validate the independence assumption adopted in \cref{sec:cond_pdf_netload}, which enables the convolution of forecast error distributions.
\cref{tab:RVs-dependencies} reports the pairwise correlation coefficients and normalized mutual information (NMI) values for the historical errors of load, wind, and solar across the NYISO zones.
The correlation values are consistently close to zero (maximum magnitude $0.09$, average $-0.0152$), indicating negligible linear dependence. Similarly, NMI values remain low (maximum $0.153$, average $0.022$), suggesting weak nonlinear dependencies. These results support the independence approximation adopted in \cref{sec:cond_pdf_netload}. Undefined values (--) correspond to zones without wind or solar resources, where the pairwise statistics cannot be computed.
%
\begin{table}[ht]
     \centering
     \caption{Pairwise Correlation and NMI Between Load, Wind, and Solar Forecast Errors Across NYISO Zones}
     \label{tab:RVs-dependencies}
     \resizebox{\columnwidth}{!}{\large
          \begin{tabular}{cccccccc}
               \toprule
                                                             & \multicolumn{3}{c}{\textbf{Correlation}}      & \multicolumn{3}{c}{\textbf{Norm. Mutual Information}}                                  \\
               \cmidrule(lr){2-4}\cmidrule(lr){5-7}
               \textbf{Area}                                 &
               $\left(\varepsilon^\ell,\varepsilon^w\right)$ & $\left(\varepsilon^\ell,\varepsilon^s\right)$ & $\left(\varepsilon^w,\varepsilon^s\right)$            &
               $\left(\varepsilon^\ell,\varepsilon^w\right)$ & $\left(\varepsilon^\ell,\varepsilon^s\right)$ & $\left(\varepsilon^w,\varepsilon^s\right)$                                             \\
               \midrule
               A                                             & -0.037                                        & 0.002                                                 & -0.070 & 0.002 & 0.029 & 0.017 \\
               B                                             & -0.063                                        & 0.090                                                 & -0.038 & 0.004 & 0.030 & 0.032 \\
               C                                             & -0.047                                        & 0.066                                                 & -0.014 & 0.002 & 0.024 & 0.015 \\
               D                                             & -0.071                                        & 0.042                                                 & 0.006  & 0.005 & 0.153 & 0.018 \\
               E                                             & -0.052                                        & 0.048                                                 & -0.051 & 0.006 & 0.019 & 0.017 \\
               F                                             & -0.002                                        & 0.080                                                 & -0.043 & 0.004 & 0.034 & 0.021 \\
               G                                             & -0.026                                        & 0.063                                                 & -0.062 & 0.005 & 0.039 & 0.035 \\
               H                                             & --                                            & --                                                    & --     & --    & --    & --    \\
               I                                             & -0.029                                        & --                                                    & --     & 0.005 & --    & --    \\
               J                                             & --                                            & --                                                    & --     & --    & --    & --    \\
               K                                             & -0.085                                        & -0.008                                                & -0.090 & 0.004 & 0.032 & 0.071 \\
               $\Sigma$                                      & -0.050                                        & 0.053                                                 & -0.038 & 0.003 & 0.004 & 0.003 \\
               \midrule
               \multicolumn{7}{l}{\normalsize Undefined values (--) indicate zones without wind or solar generation.}                                                                                 \\
               \bottomrule
          \end{tabular}%
     }
\end{table}

Two metrics are used throughout the experiments to assess the performance of the reserve sizing methods:
\begin{enumerate}[(i)]
     \item Coverage: The percentage of deviations captured within the estimated reserve intervals. An ideal method achieves coverage close to the specified target (e.g., 90\%).
     \item Normalized Average Reserve Requirement (NARR):
           To enable comparison across areas of different sizes, we define the Normalized Average Reserve Requirement (NARR).
           Let the Average Reserve Requirement (ARR) of method or configuration $m$ be
           \begin{equation}
                \text{ARR}^{(m)}
                = \frac{1}{n}\sum_{i=1}^n
                r^{\textrm{up},\left(m\right)}_i + r^{\textrm{down},\left(m\right)}_i,
           \end{equation}
           where $r^{\textrm{up},\left(m\right)}_i$ and $r^{\textrm{down},\left(m\right)}_i$ are the upward and downward reserve requirements at time $i$.

           The normalized metric is then defined as
           \begin{equation}
                \textrm{NARR}^{\left(m\right)} = \frac{\textrm{ARR}^{\left(m\right)}}
                {\textrm{ARR}^{\left(\textrm{ref}\right)}},
           \end{equation}
           where $\textrm{ARR}^{\left(\textrm{ref}\right)}$ is the reference value used for normalization. This reference can be chosen in different ways: for instance, the largest ARR across all methods (to emphasize efficiency relative to the most conservative approach), or a specific benchmark method, such as the proposed framework (to highlight relative performance against the method of interest).
           By construction, $\textrm{NARR}$ provides a dimensionless comparison of reserve efficiency, with values below one indicating smaller average reserve requirements than the chosen reference.
           NARR measures reserve-volume efficiency and is not converted to monetary value here, as such a conversion would require additional assumptions (e.g., reserve prices, opportunity costs, or a production-cost simulation).
\end{enumerate}

Together, coverage and NARR provide complementary insights: coverage evaluates reliability, while NARR evaluates efficiency.

\subsection{Ablation Study of Modeling Choices for Reserve Sizing}\label{sec:ablation_study}
For deterministic forecasts, conditional error distributions must be built from historical deviations using selected explanatory variables and binning strategies. This ablation evaluates both choices: explanatory-variable selection is tested as a main modeling decision, while the binning comparison serves as a sensitivity check for the local density-estimation step. For probabilistic forecasts, the conditional distribution is provided directly for each forecast time, as discussed in \cref{sec:prob_forecast}.

The binning strategies considered are:
\begin{itemize}
     \item $\textrm{B}_0$: no binning (all data pooled together),
     \item $\textrm{B}_1$: fixed-size bins,
     \item $\textrm{B}_2$: variable-size bins, and
     \item $\textrm{B}_3$: $k$-nearest neighbor ($k$-NN) bins, corresponding to the binning strategy used in the proposed configuration.
\end{itemize}
Note that $\textrm{B}_0$--$\textrm{B}_2$ serve as comparison cases for evaluating alternatives to $k$-NN binning.

\begin{figure}[ht]
     \centering
     \includegraphics[width=\columnwidth]{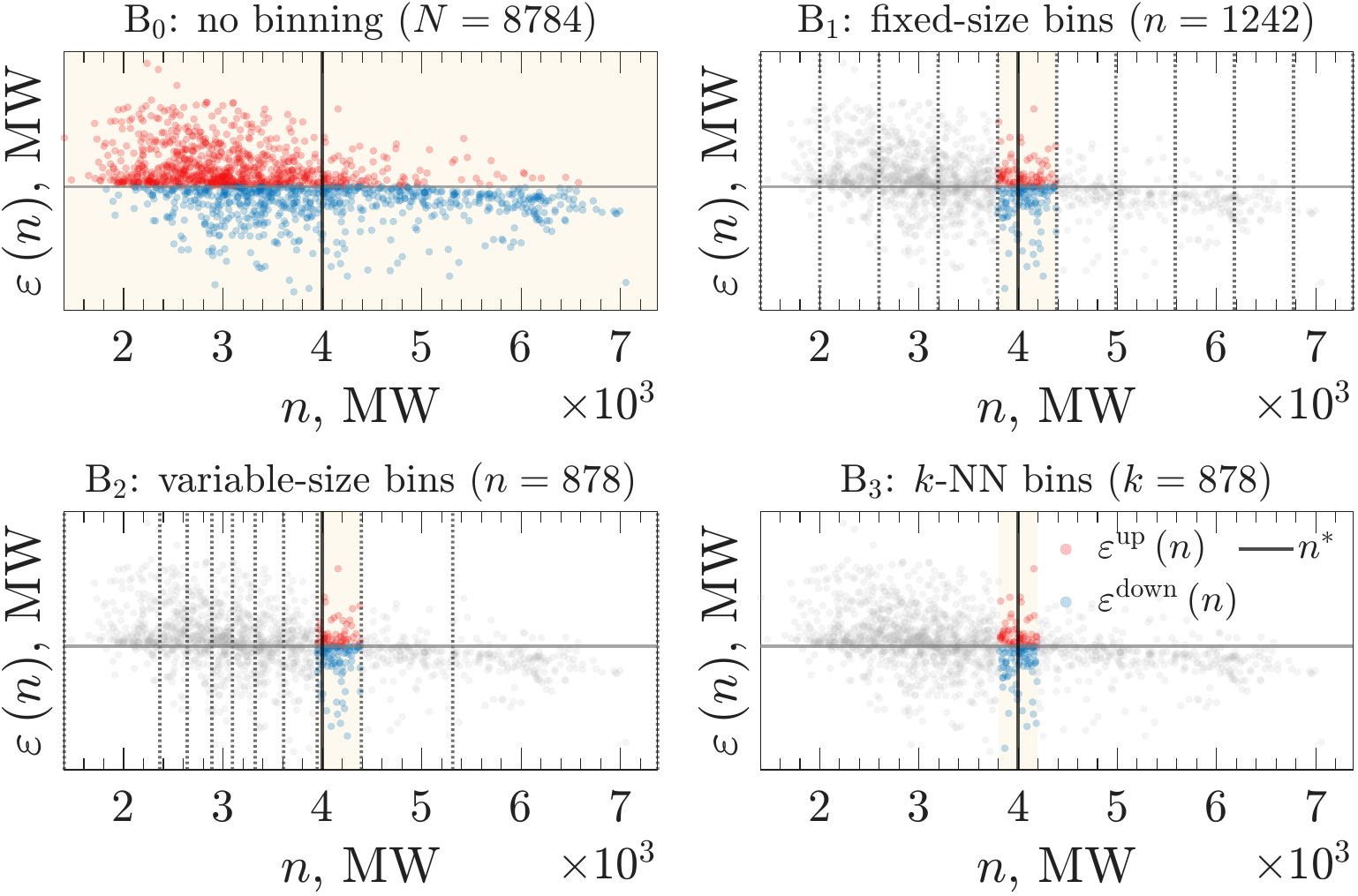}
     \caption{The four binning strategies applied to illustrative net-load error data at an evaluation point $n^{*}$ (black line): the highlighted samples are those selected to estimate $f\left(\varepsilon \mid n^{*}\right)$.}
     \label{fig:binning_strategies}
\end{figure}
\cref{fig:binning_strategies} shows how the four strategies select samples from illustrative net-load error data for a common evaluation point $n^{*}=4{,}000$~MW. Each panel plots the net-load error $\varepsilon$ against the explanatory variable $n$ and highlights the samples retained to estimate $f\left(\varepsilon \mid n^{*}\right)$. Strategy $\textrm{B}_0$ pools all observations, ignoring $n^{*}$; $\textrm{B}_1$ retains the fixed-width bin containing $n^{*}$, so the number of samples depends on the local data density; $\textrm{B}_2$ retains the equal-count bin containing $n^{*}$, so the bin width adapts while the sample count stays fixed; and $\textrm{B}_3$ retains the $k$ nearest samples to $n^{*}$. Progressing from $\textrm{B}_0$ to $\textrm{B}_3$ narrows the conditioning set to a local, adaptive neighborhood around the forecasted value, motivating the use of $k$-NN in the proposed method.
For explanatory variable selection, we compare two approaches:
\begin{itemize}
     \item Correlation: selection based on Pearson correlation, and
     \item NMI: selection based on normalized mutual information, normalized to $\left[0,1\right]$.
\end{itemize}

Two performance metrics are used: (i) coverage, fixed at a 90\% target, and (ii) the Normalized Average Reserve Requirement (NARR), as defined in \cref{sec:dataset_setup}.
In this subsection, NARR is normalized with respect to the largest average reserve requirement observed across all configurations, ensuring fair comparisons of efficiency. An ideal configuration achieves coverage close to the 90\% target while minimizing NARR, i.e., high reliability with short reserve intervals.

\begin{figure}[htb]
     \centering
     \begin{subfigure}[t]{1.65in}
          \centering
          \includegraphics[width=\textwidth]{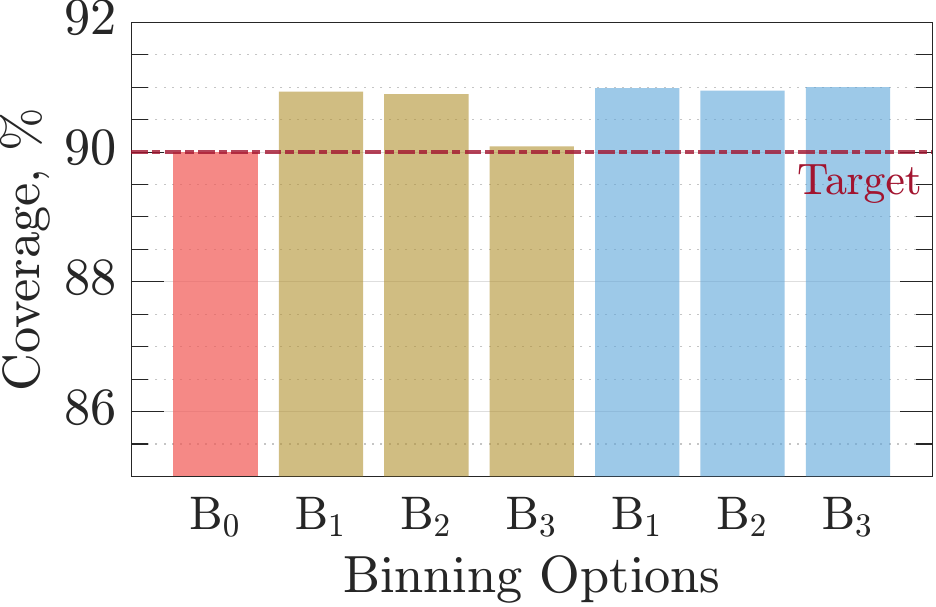}
          \caption{Average Coverage.}
          \label{fig:ablation_coverage}
     \end{subfigure}
     \hfill
     \begin{subfigure}[t]{1.65in}
          \centering
          \includegraphics[width=\textwidth]{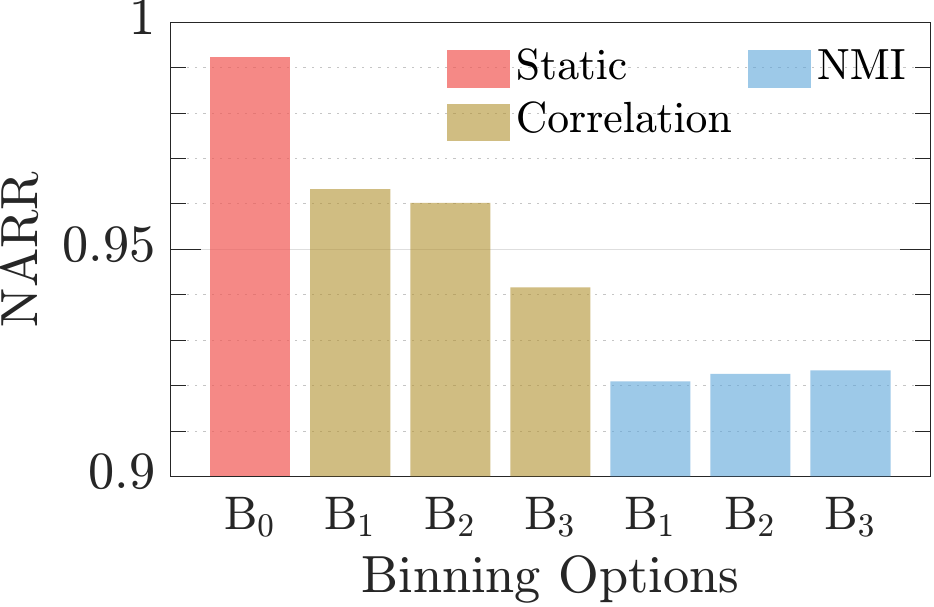}
          \caption{Average NARR.}
          \label{fig:ablation_length}
     \end{subfigure}
     %
     \caption{Average coverage and average NARR across all NYISO zones.}
     \label{fig:ablation_study}
\end{figure}

Each configuration is evaluated across the eleven NYISO zones (A–K) and the system aggregate ($\Sigma$). Results, in terms of average coverage and average NARR across all areas, are summarized in \cref{fig:ablation_study}.

As shown in \cref{fig:ablation_coverage}, all model configurations attain the desired 90\% coverage, consistent with the construction of the confidence intervals in \cref{eq:CI_quantile}. The main separation appears in reserve efficiency: $\textrm{B}_0$ produces the widest intervals because all data are pooled, while NMI-based configurations produce lower NARR than correlation-based configurations across the tested binning options. Among the NMI-based configurations, $\textrm{B}_1$--$\textrm{B}_3$ produce comparable average NARR values, so the ablation does not show a meaningful efficiency difference among these binning options for this dataset. We use $\textrm{B}_3$ ($k$-NN) as the baseline because it is consistent with the local conditional-density estimation used in the proposed method: the neighborhood adapts to the density of nearby samples rather than relying on fixed bin boundaries. The proposed deterministic configuration therefore combines NMI for explanatory variable selection with $k$-NN binning in the subsequent comparisons.

\subsection{Reserve Sizing Method Benchmark}
This subsection evaluates the selected deterministic configuration from \cref{sec:ablation_study}, where conditional error distributions are derived from historical deviations. Probabilistic forecasts are not used in this benchmark; they are assessed separately in the next subsection. The performance of the proposed method is compared against two benchmarks:
\begin{enumerate}[(i)]
     \item \textit{Static Reserves}: Upward and downward reserves are estimated by computing a desired quantile of the forecast error distribution using the complete dataset (i.e., a single bin).  This yields fixed requirements for any forecast and represents a simple non-adaptive baseline, which we denote as \textit{Static}. It is used only as a comparison case and is not intended to represent NYISO's current reserve requirements.
     \item \textit{Dynamic reserves under Gaussian Assumption}:
           Reserves are calculated assuming that forecast errors follow a Gaussian distribution. In earlier studies \cite{holttinen2012methodologies,krad2015quantifying}, this approach is implemented using fixed-size bins ($\textrm{B}_1$) and correlation for explanatory variable selection. This method is referred to as the \textit{Gaussian} method.
\end{enumerate}
To ensure statistical robustness, the dataset is randomly partitioned by day 100 times. In each of 100 independent partitions, 70\% of the days (255) are used to construct the conditional error distributions, and the remaining 30\% (110) are reserved for evaluation. Performance is always assessed on the evaluation set, which is not used to build the distributions. The final results are reported as averages across all partitions, providing stable estimates and reducing sensitivity to any single split.

\subsubsection{Coverage and Reserve Efficiency}
\begin{figure}[htb]
     \centering
     \begin{subfigure}[t]{1.72in}
          \centering
          \includegraphics[width=\textwidth]{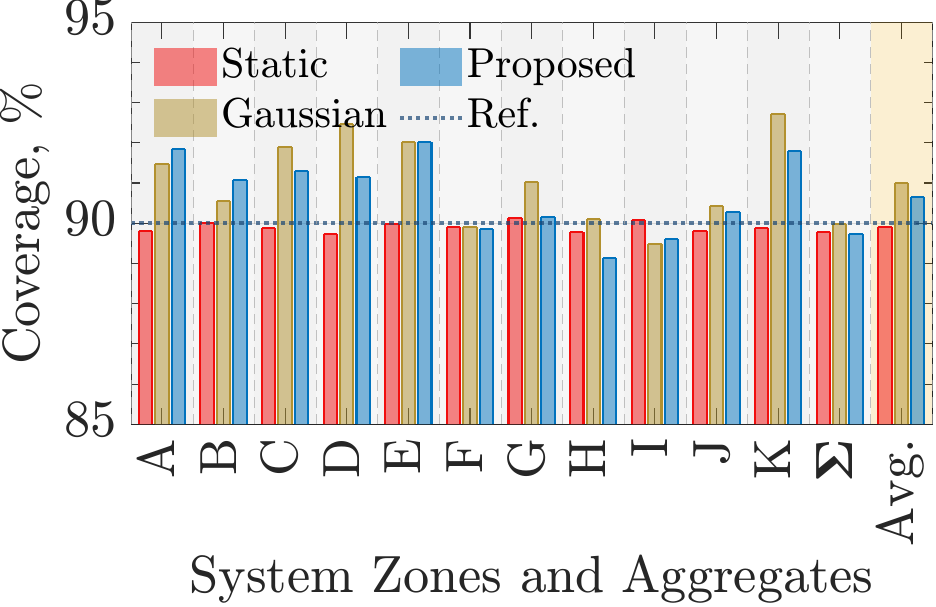}
          \caption{Coverage performance of the Static, Gaussian, and Proposed methods, with the 90\% target shown for reference.}
          \label{fig:coverage_comparison}
     \end{subfigure}
     \hfill
     \begin{subfigure}[t]{1.72in}
          \centering
          \includegraphics[width=\textwidth]{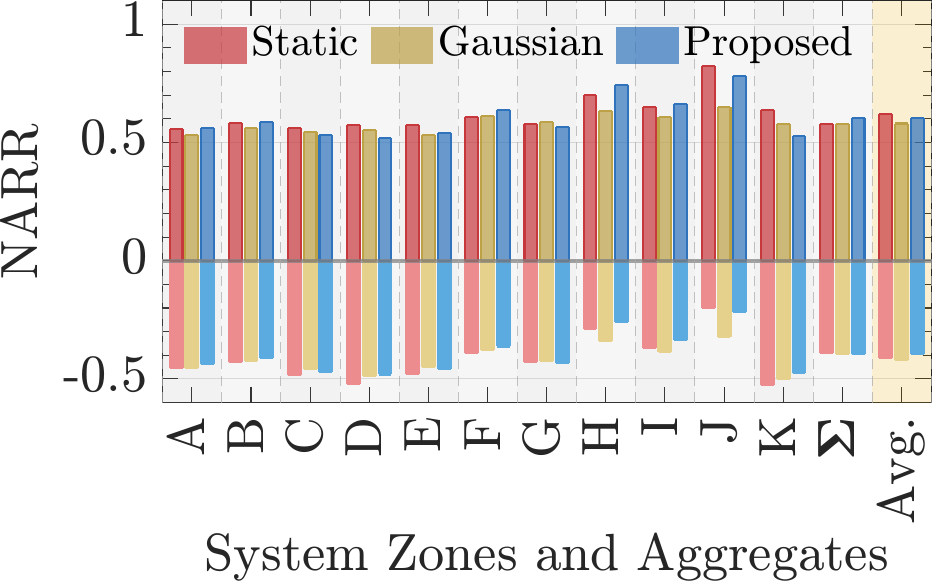}
          \caption{Normalized Average Reserve Requirement (NARR) for upward (positive) and downward (negative) reserves.}
          \label{fig:NARR_comparison}
     \end{subfigure}
     %
     \caption{Coverage and reserve efficiency across NYISO areas.}
     \label{fig:coverage_vs_NARR}
\end{figure}

Two metrics are used to evaluate the reserve sizing methods:
(i) coverage, which measures how often forecast deviations fall within the sized reserves, and
(ii) reserve efficiency, assessed through the \emph{Normalized Average Reserve Requirement} (NARR), computed separately for upward and downward reserves.
In \cref{fig:NARR_comparison}, each bar shows the required upward (above zero) and downward (below zero) normalized reserves for each NYISO area and the system aggregate.

The average coverages across all NYISO zones (A–K) and the system aggregate are shown in~\cref{fig:coverage_comparison}. The methods generally meet or exceed the 90\% target, although the Proposed method falls slightly below the target in zones H and I.
The Gaussian method tends to over-cover because its symmetric parametric form produces prediction intervals that are wider than needed in some operating conditions.
In several other areas, the Proposed method exhibits slight over-coverage. This occurs for a different reason: in regions where data are sparse, the $k$-NN procedure expands the local neighborhood to obtain a stable conditional density estimate, which can lead to conservative interval bounds near the edges of the distribution.

Reserve efficiency, shown in~\cref{fig:NARR_comparison}, highlights the differences in reserve sizing across methods. All three methods exhibit consistent asymmetry between upward and downward requirements across nearly all zones.
While directional NARR values alone cannot rigorously establish skewness, the systematic imbalance between positive and negative deviations across areas and methods strongly suggests that the empirical error distribution is biased and asymmetric around zero, likely skewed, with unequal dispersion on either side of zero.
%

Since the Gaussian method assumes a symmetric, thin-tailed error distribution, the asymmetry observed in its NARR values indicates a structural mismatch with the empirical data: a symmetric model cannot characterize the directional differences present in real forecast errors, producing NARR patterns that may resemble those of the empirical method but arise from incorrect distributional assumptions.

Risk measures such as CVaR depend on the shape and thickness of the tails, so if a model misrepresents one tail, the resulting reserves will not match the actual operational risk. Under the Gaussian assumption, the imposed symmetry and thin tails underestimate the severity of large deviations, resulting in \emph{optimistic risk estimates}, whereas the Proposed method retains the directional structure of the empirical errors and better reflects the actual tail behavior. As a result, even when total reserve volumes appear similar, the models' assumptions can lead to discrepancies in estimated risk, as the next section shows.

\subsubsection{Quantitative Risk Comparison}
The directional NARR asymmetries indicate that the empirical error distribution exhibits uneven and heavier tails. Because these tails govern both the probability and the severity of reserve-violating events, reserve magnitudes alone cannot reveal how well each method characterizes operational risk; a meaningful comparison requires examining how accurately each method models the tail behavior.

To quantify these differences, we evaluate the operational risk implied by the Gaussian method relative to that obtained from the empirical conditional distribution.
For each NYISO area and for both shortfall (upward) and longfall (downward) directions, we compute the percentage error in risk when the Gaussian model is used instead of the empirical benchmark.
Positive values indicate that the Gaussian method \emph{underestimates} true risk, whereas negative values indicate that it \emph{overestimates} it.

\begin{figure}[ht]
     \centering
     \includegraphics[width=1.8in]{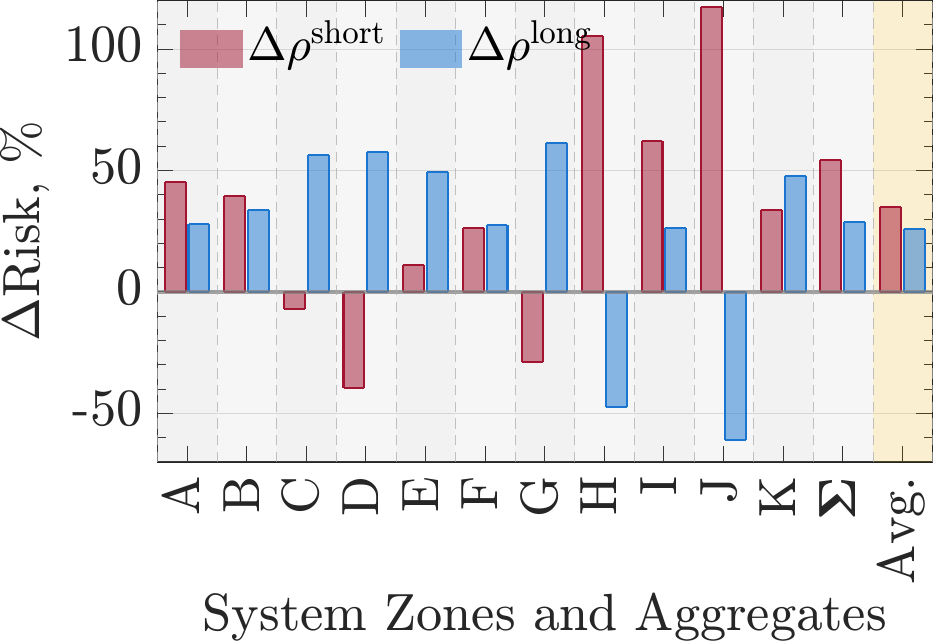}
     \caption{Percentage change in risk under the Gaussian assumption relative to the empirical distribution. Positive values indicate that the Gaussian model \emph{underestimates} the true risk, while negative values indicate that it \emph{overestimates} it.}
     \label{fig:delta_risk}
\end{figure}

The results in \cref{fig:delta_risk} show a consistent pattern: in most areas and in both directions, the Gaussian method underestimates risk, often by substantial margins.
This behavior follows directly from its imposed symmetry and thin-tailed structure, which assigns too little probability mass to large deviations, particularly in directions where the empirical distribution is biased or heavy-tailed.
Occasional negative values correspond to instances in which the Gaussian model inadvertently oversizes reserves, but these cases are fewer and generally smaller in magnitude.

Overall, this analysis demonstrates that the Gaussian assumption systematically mischaracterizes tail behavior and produces \emph{overly optimistic risk estimates}.
%

\subsection{Risk Assessment with Deterministic and Probabilistic Forecast Inputs}
As described in \cref{sec:cond_pdf_netload}, the proposed framework is flexible and can accommodate both deterministic and probabilistic forecasts. Unlike the benchmark in the previous subsection, this analysis evaluates how reserve and risk estimates change when historical-error models are replaced by probabilistic forecasts for some or all resources. To assess this effect, we consider the five configurations listed in \cref{tab:prob-det-cases}, which range from fully probabilistic to fully deterministic forecasts.

First, the day in the dataset with the highest forecast uncertainty under the probabilistic configuration is identified. Then the reserve requirements are determined across the five configurations, and the resulting risks are compared.

\subsubsection{Identification of the Day with the Largest Forecast Uncertainty}
Forecast uncertainty can be quantified rigorously and interpretably using \emph{entropy}. Entropy, $h(\cdot)$, measures the dispersion of a random variable~\cite{cover2006elements}: low entropy indicates that the distribution is concentrated, while high entropy indicates it is widely spread. Applied to probabilistic forecasts:
$
     h\left(X_{t,d}\right) = -\int
     f_{t,d}\left(x\right) \log f_{t,d}\left(x\right) dx,
$
where $f_{t,d}\left(x\right)$ is the probability density of the forecast distribution at time $t$ on day $d$. The daily aggregate entropy is then defined as
\begin{equation}\label{eq:entropy_day}
     \varphi_d = \sum_{t\in\cT} h\left(X_{t,d}\right), \qquad d\in\cD.
\end{equation}

\begin{figure}[ht]
     \centering
     \includegraphics[width=3.3in]{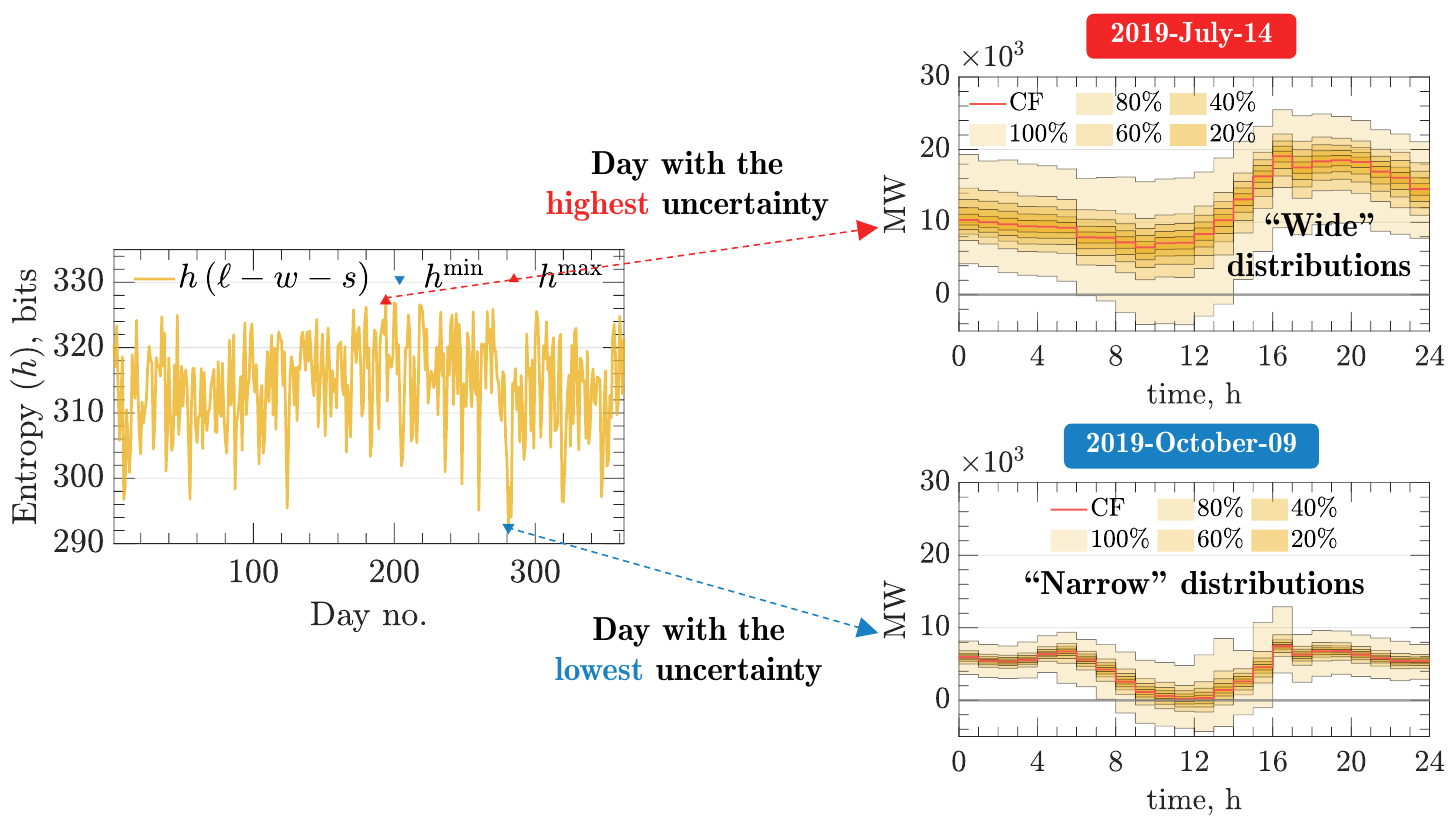}
     \caption{Aggregated Entropy per Day}
     \label{fig:agg_entropy_day}
\end{figure}

Using \cref{eq:entropy_day}, aggregated entropy is computed for the net load forecasts in the dataset. Results are shown in \cref{fig:agg_entropy_day}, where the least uncertain day is Oct. 9\textsuperscript{th} and the most uncertain is July 14\textsuperscript{th}. On July 14\textsuperscript{th}, the probabilistic forecasts exhibit wide prediction bands and skewed, heavy-tailed distributions. Such features reflect meteorological conditions with high uncertainty, leading to broad, asymmetric probability distributions of the net load.

\subsubsection{Risk Assessment Using Probabilistic Forecasts}
July 14\textsuperscript{th} is chosen as the case study because it exhibits the highest forecast uncertainty (\cref{fig:agg_entropy_day}), providing a stringent test of the framework, since reserve requirements and risk exposures are most sensitive under highly uncertain conditions.

The five forecast configurations listed in \cref{tab:prob-det-cases} are examined.
Case (i), which employs three probabilistic forecasts $\fcast{\ProbF}{\ProbF}{\ProbF}$, is referred to as the probabilistic forecast configuration and serves as the reference case, since it uses day-ahead uncertainty distributions for load, wind, and solar on the analyzed day and forecast horizon.
Cases (ii)-(iv) are mixed configurations, and case (v) is the fully deterministic configuration. Their associated risks are evaluated relative to this reference case.
In all experiments, the risk ceilings are set symmetrically, $\gamma^{\text{short}} = \gamma^{\text{long}} = 50$~MW, corresponding to the average reference-case risk at a 90\% confidence level.

\begin{table}[ht]
     \centering
     \caption{Cases used to assess the impact of probabilistic forecasts on reserve determination.}
     \label{tab:prob-det-cases}
     \begin{tabular}{@{} cccc c @{}}
          \toprule
                & \multicolumn{3}{c}{\textbf{Forecast type}} &                                                                        \\
          \cmidrule(lr){2-4}
          \textbf{Case}
                & $\boldsymbol{\ell}$                        & $\boldsymbol{w}$ & $\boldsymbol{s}$ &
          \textbf{Notation}                                                                                                           \\
          \midrule
          (i)   & \ProbF                                     & \ProbF           & \ProbF           & $\fcast{\ProbF}{\ProbF}{\ProbF}$ \\
          (ii)  & \ProbF                                     & \Det             & \Det             & $\fcast{\ProbF}{\Det}{\Det}$     \\
          (iii) & \Det                                       & \ProbF           & \Det             & $\fcast{\Det}{\ProbF}{\Det}$     \\
          (iv)  & \Det                                       & \Det             & \ProbF           & $\fcast{\Det}{\Det}{\ProbF}$     \\
          (v)   & \Det                                       & \Det             & \Det             & $\fcast{\Det}{\Det}{\Det}$       \\
          \bottomrule
     \end{tabular}

     \vspace{2pt}
     {\small\ProbF{}: Probabilistic forecast; \Det{}: Deterministic forecast}
\end{table}

\begin{figure}[htb]
     \centering
     \begin{subfigure}[t]{1.72in}
          \centering
          \includegraphics[width=\textwidth]{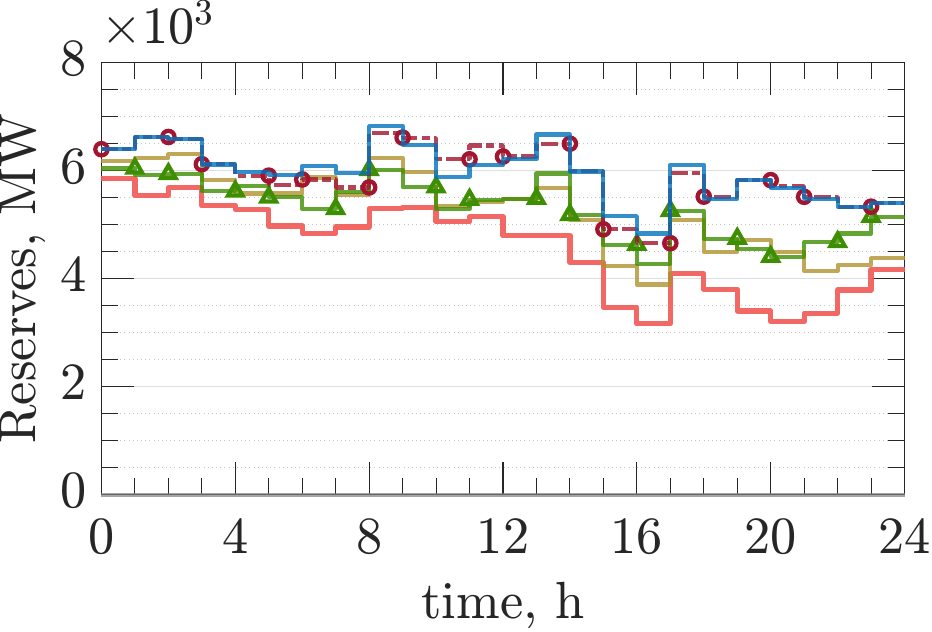}
          \caption{Upward Reserves.}
          \label{fig:reserve_up_comparison}
     \end{subfigure}
     \hfill
     \begin{subfigure}[t]{1.72in}
          \centering
          \includegraphics[width=\textwidth]{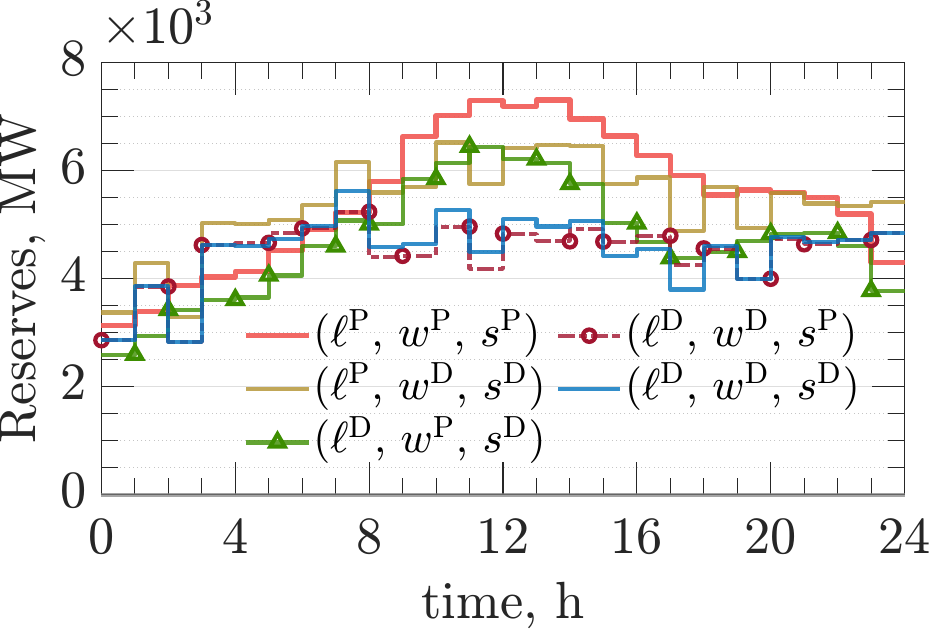}
          \caption{Downward Reserves.}
          \label{fig:reserve_down_comparison}
     \end{subfigure}
     \\[.7em] 
     \begin{subfigure}[t]{1.72in}
          \centering
          \includegraphics[width=\textwidth]{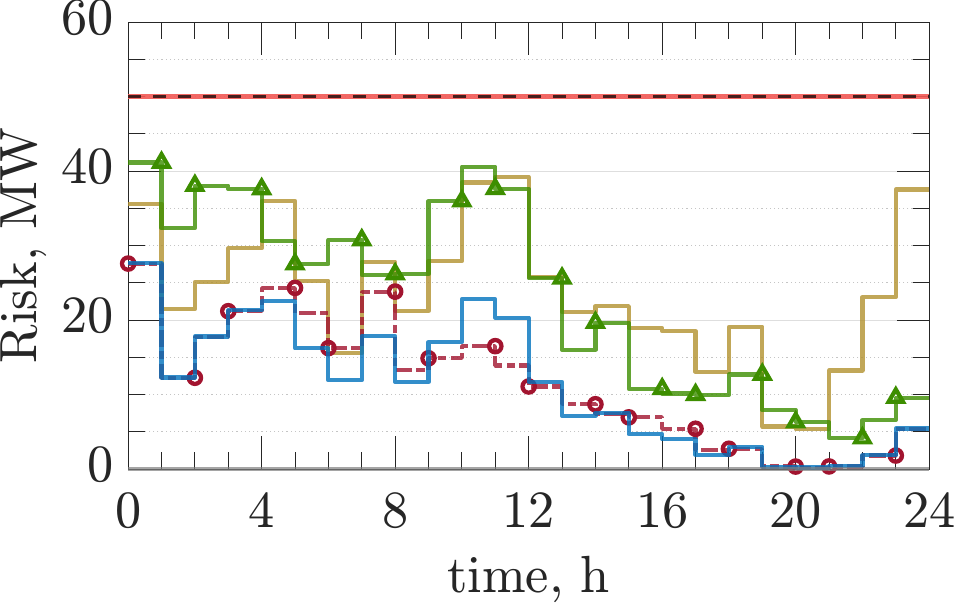}
          \caption{Risk of being Short.}
          \label{fig:risk_short_comparison}
     \end{subfigure}
     \hfill
     \begin{subfigure}[t]{1.72in}
          \centering
          \includegraphics[width=\textwidth]{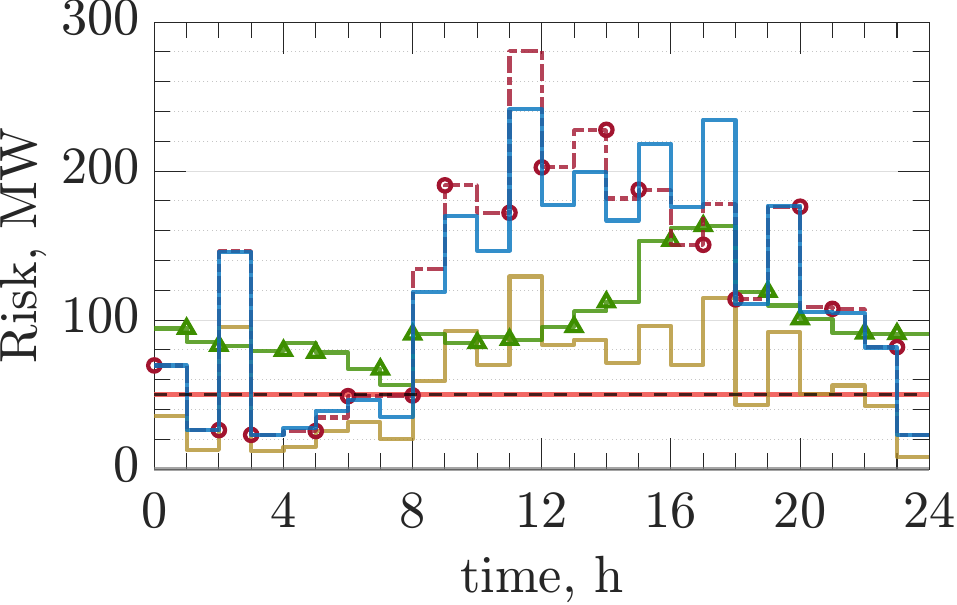}
          \caption{Risk of being Long.}
          \label{fig:risk_long_comparison}
     \end{subfigure}
     %
     \caption{Reserve and risk comparison; colors correspond to \cref{tab:prob-det-cases}.}
     \label{fig:reserve_risk_comparison}
\end{figure}

\cref{fig:reserve_risk_comparison} shows the reserve requirements and associated risks for the five forecast configurations.
The probabilistic forecast configuration estimates reserves directly from day-ahead uncertainty distributions tailored to the forecasted day's meteorological conditions, whereas deterministic configurations derive reserves from historical deviations across diverse operating conditions.
For this dataset, the historical-based distributions exhibit greater overall uncertainty than the probabilistic forecasts. Notably, the uncertainty behaves differently across directions: the upward uncertainty is consistently smaller in the probabilistic forecasts than in the historical-based distributions, whereas the downward uncertainty tends to be larger. These contrasting behaviors directly influence reserve sizing and risk estimation, as demonstrated in the following experiments. The main findings are summarized below.

\begin{itemize}
     \item Upward reserves (\cref{fig:reserve_up_comparison}):
           The probabilistic forecast configuration results in lower upward reserves, consistent with its narrower uncertainty distributions, whereas deterministic configurations are derived from aggregated past deviations and therefore produce broader, more conservative upward reserves.
     \item Downward reserves (\cref{fig:reserve_down_comparison}):
           During the first eight hours, the probabilistic configuration sets smaller downward reserves, indicating lower expected deviations. After approximately eight hours, the relationship reverses as the probabilistic configuration captures the higher likelihood of longfall events later in the day.
     \item Shortfall risks (\cref{fig:risk_short_comparison}):
           Deterministic configurations, with larger upward reserves, exhibit lower shortfall risk but at the cost of excessive reserve sizing. The probabilistic configuration keeps shortfall risks within acceptable limits while avoiding unnecessary oversizing.
     \item Longfall risks (\cref{fig:risk_long_comparison}):
           On the downward side, deterministic configurations have higher longfall risk during several hours, as smaller reserves fail to capture periods of higher uncertainty. The probabilistic configuration mitigates these risks by increasing downward reserves when the likelihood of longfall events rises.
\end{itemize}

Overall, these results support using reliable probabilistic forecasts when available, because they provide day-specific uncertainty information that historical deterministic errors cannot fully capture \cite{NCE2023}. For this study day, the fully probabilistic configuration best represents expected operating conditions, while mixed configurations use probabilistic forecasts where available and historical-error models otherwise.
%
%

\section{Conclusion}\label{sec:conclusion}
This paper presents a source-agnostic framework for dynamic flexibility reserve sizing. The framework constructs conditional error distributions for load, wind, and solar, combines them into a conditional net-load error distribution, and translates it into upward and downward reserve requirements using probabilistic and risk-based criteria. These distributions can be derived from historical forecast deviations or probabilistic forecasts, allowing the method to use empirical error patterns and day-specific operational uncertainty information when available.

Numerical experiments on an NYISO-based synthetic dataset show that the framework improves reserve-volume efficiency over static benchmarks while maintaining reliability. Compared with Gaussian-based parametric methods, it avoids the systematic underestimation of tail risk caused by symmetric, light-tailed error assumptions. The ablation study shows that NMI-based explanatory-variable selection improves efficiency relative to correlation-based selection, while $k$-NN binning provides the local density-estimation structure used in the proposed deterministic configuration. The deterministic/probabilistic comparison indicates that probabilistic forecasts are preferred when available and reliable, since historical information alone cannot capture day-specific weather-driven uncertainty.

The resulting reserve requirements reflect directional differences between upward and downward deviations while explicitly quantifying the risk of uncovered deviations. Because these requirements are built directly from the resource uncertainty distributions, the framework remains transparent and interpretable: operators can trace each reserve value back to the underlying distributions and conditioning variables, understanding why a given reserve level is recommended. The framework is computationally tractable and implemented in EPRI's DynADOR tool for reserve scheduling. Moreover, by yielding explicit reserve requirements, it integrates seamlessly into existing production-cost models as deterministic reserve constraints, avoiding the added computational complexity of scenario-based stochastic optimization. Future work will relax independence assumptions among resources, incorporate spatiotemporal correlations, quantify economic impacts through production-cost simulations, and apply the framework to stochastic unit commitment and market-based reserve co-optimization. Machine-learning methods are also a promising alternative for reserve sizing, and integrating them in a safe and transparent manner is another direction for future work.

\bibliographystyle{IEEEtran}
\bibliography{utils/references}

\end{document}